# Quadratic soliton mode-locked degenerate optical parametric oscillator

MINGMING NIE,* AND SHU-WEI HUANG*

*Department of Electrical, Computer & Energy Engineering, University of Colorado Boulder, Boulder, CO 80309, USA*
*Corresponding author: Mingming.Nie@colorado.edu, ShuWei.Huang@colorado.edu*

**By identifying the similarities between the coupled-wave equations and the parametrically driven nonlinear Schrödinger equation, we unveil the existence condition of the quadratic soliton mode-locked degenerate optical parametric oscillator in the previously unexplored parameter space of near-zero group velocity mismatch. We study the nature of the quadratic solitons and divide their dynamics into two distinctive branches depending on the system parameters. We find the nonlinear interaction between the resonant pump and signal results in phenomena that resemble the dispersive two-photon absorption and the dispersive Kerr effect. Origin of the quadratic soliton perturbation is identified and strategy to mitigate its detrimental effect is developed. Terahertz comb bandwidth and femtosecond pulse duration are attainable in an example periodically poled lithium niobate waveguide resonator in the short-wave infrared and an example orientation-patterned gallium arsenide free-space cavity in the long-wave infrared. The quadratic soliton mode-locking principle can be extended to other material platforms, making it a competitive ultrashort pulse and broadband comb source architecture at the mid-infrared.**

## 1. INTRODUCTION

Mode-locked laser (MLL) and optical frequency comb (OFC) have been the cornerstones and key enabling technologies for many scientific breakthroughs in precision frequency metrology, ultrastable time keeping, nonlinear bioimaging, atmospheric remote sensing, and more. Decades of detailed investigation and careful optimization have led to routine generation of few-cycle pulses directly from Ti:sapphire laser oscillators, or by series of nonlinear optical wave mixings [1-3]. Extension of MLL and OFC technologies from the traditional near-infrared regime to new spectral regions is likely to trigger new research opportunities in both science and technology. The mid-infrared (MIR) spectral region especially attracts great attention due to the presence of strong rovibrational transitions in many molecules, making possible detailed molecular fingerprinting.

The choice of gain media and mode-locking techniques that are suitable for MIR mode-locked lasers, however, is quite limited and laser sources at wavelengths beyond 3 μm remain scarce [4, 5]. On the other hand, optical parametric oscillator (OPO) is intrinsically broadband and tunable as it does not depend on atomic or molecular resonances. Synchronous pumping scheme, in which the circulating OPO signal is periodically amplified by a MLL that is synchronized to the OPO cavity, has been demonstrated effective and synchronously pumped OPOs based on periodically poled lithium niobate (PPLN) [6, 7] and orientation-patterned gallium arsenide (OP-GaAs) [8, 9] have been successfully implemented as the viable MIR femtosecond light sources.

However, synchronously pumped OPOs require additional MLLs and associated synchronization electronics, thus generally resulting in relatively high complexity, large footprint, and high cost for such OPOs. To address the issues, techniques to mode-lock continuous-wave (cw) pumped OPO have been investigated and developed. Early efforts in this research direction focused on active mode-locking with intracavity electro-optic modulator and acousto-optic modulator [10-15]. The first attempts towards the passively mode-locked OPO were reported in 2013 and 2014 where the intracavity phase mismatched second harmonic generation (SHG) was utilized [16, 17]. Later demonstration and analysis showed that phase matched SHG can also promote the mode-locking action [18-20]. A recent study further suggests that group velocity mismatch (GVM) between the fundamental field (FF) and the second harmonic (SH) results in the domain wall locking phenomena that facilitates the formation of mode-locked pulses [20]. However, none of these prior reports demonstrate a fully mode-locked OPO with stable femtosecond pulses and coherent comb spectra. A question then arises whether high-quality passively mode-locked OPO can ever exist in cw-pumped quadratic nonlinear cavities, similar to the dissipative Kerr soliton and Kerr frequency comb generation in cw-pumped cubic nonlinear cavities [21, 22].

In this letter, we unveil the existence condition of high-quality passively mode-locked degenerate OPO (DOPO) in the previously unexplored parameter space of near-zero GVM. We first study the special case of zero GVM and then expand the analysis to include the effect of temporal walk-off. We identify the similarities between the coupled-wave equations describing the DOPO dynamics and the parametrically driven nonlinear Schrödinger equation (NLSE) [23, 24]. Such an analogy helps elucidate the underlying mechanism of the quadratic soliton dynamics. We study the nature of the quadratic soliton mode-locked DOPO and divide the quadratic soliton dynamics into two distinctive branches depending on the system parameters. We find the nonlinear interaction between the resonant pump and signal results in phenomena that resemble the dispersive two-photon absorption (TPA) and the dispersive Kerr effect. We investigate their perturbation to the quadratic soliton and develop the strategy to avoid their detrimental effects. Finally, we give two realistic examples demonstrating terahertz comb bandwidth and femtosecond pulse duration are attainable in both a PPLN waveguide microresonator and an OP-GaAs free-space cavity.

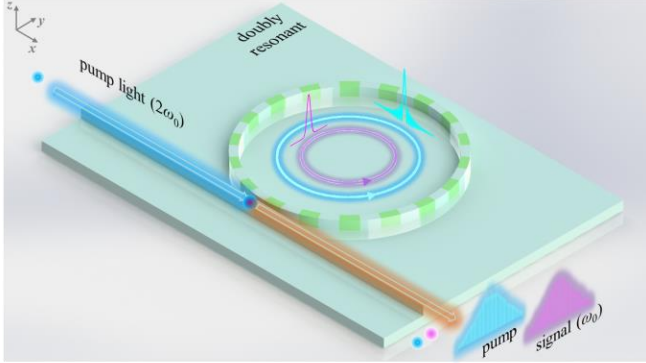

Fig. 1. Schematic of the quadratic soliton mode-locked DOPO. With the proper pump and cavity parameters, ultrashort pulses and broadband combs can be generated in cw-pumped doubly resonant cavities with quadratic nonlinearities.

## 2. THEORETICAL ANALYSIS AND NUMERICAL RESULTS

The field evolution in the retarded time frame through a cw-pumped DOPO (Fig. 1) obeys the coupled equations:

$$\frac{\partial A}{\partial z} = \left[-\frac{\alpha_{c1}}{2} - i\frac{k_1^{"}}{2}\frac{\partial^2}{\partial \tau^2}\right]A + i\kappa BA^* e^{-i\Delta kz}, \quad (1)$$

$$\frac{\partial B}{\partial z} = \left[-\frac{\alpha_{c2}}{2} - \Delta k^{'}\frac{\partial}{\partial \tau} - i\frac{k_2^{"}}{2}\frac{\partial^2}{\partial \tau^2}\right]B + i\kappa A^2 e^{i\Delta kz}, \quad (2)$$

and the boundary conditions:

$$A_{m+1}(0,\tau) = \sqrt{1-\theta_1} A_m(L,\tau) e^{-i\delta_1}, \quad (3)$$

$$B_{m+1}(0,\tau) = \sqrt{1-\theta_2} B_m(L,\tau) e^{-i\delta_2} + \sqrt{\theta_2} B_{in}, \quad (4)$$

where $A$ is the signal field envelope, $B$ is the pump field envelope, $B_{in}$ is the cw pump, $\alpha_{c1,2}$ are the propagation losses, $\Delta k$ is the wave-vector mismatch, $\Delta k^{'}$ is the group-velocity mismatch (GVM), and $k_{1,2}^{"}$ are the group-velocity dispersion (GVD) coefficients, $L$ is the nonlinear medium length, $\theta_{1,2}$ are the coupler transmission coefficients and $\delta_{1,2}$ are the pump-resonance and signal-resonance phase detuning, respectively [25]. $\kappa = \sqrt{2}\omega_0 d_{eff}/\left(A_{eff}\sqrt{c^3 n_1^2 n_2 \varepsilon_0}\right)$ is the normalized second-order nonlinearity coupling coefficient, where $d_{eff}$ is the effective second-order nonlinear coefficient, $A_{eff}$ is the effective mode area, $c$ is the speed of light, $\varepsilon_0$ is the vacuum permittivity, and $n_{1,2}$ are the refractive indices. Higher-order dispersion and nonlinearity are both neglected for simplicity.

Under the mean-field and good cavity approximations [26], Eqs. (1)-(4) can be simplified into a single mean-field equation for the signal field:

$$t_R \frac{\partial A}{\partial t} = \left(-\alpha_1 - i\delta_1 - i\frac{k_1^{"}L}{2}\frac{\partial^2}{\partial \tau^2}\right)A \\ -\left(\kappa L \text{sinc}(\xi)\right)^2 A^*\left[A^2 \otimes J(\tau)\right] + i\mu A^*, \quad (5)$$

where $t$ is the "slow time" that describes the envelope evolution over successive round-trips, $t_R$ is the roundtrip time, $\tau$ is the "fast time" that depicts the temporal profiles in the retarded time frame, $\xi = \Delta kL/2$ is the wave-vector mismatch parameter, and $\alpha_1$ is the total signal linear cavity loss. The fourth term on the right-hand side is the effective third-order nonlinearity where the nonlinear response function

$$J(\Omega) = \frac{1}{\alpha_2 + i\delta_2 - i\Delta k^{'} L\Omega - i k_2^{"} L\Omega^2/2}, \quad (6)$$

and $J(\tau) = \mathcal{F}^{-1}[J(\Omega)]$ describes the dispersion of the effective third-order nonlinearity. Here, $\Omega$ is the offset angular frequency with respect to the signal resonance frequency, $\alpha_2$ is the total pump linear cavity loss. The last term on the right-hand side, $\mu = \kappa L e^{i(\psi-\xi)}\text{sinc}(\xi)\sqrt{\theta_2}B_{in}/\sqrt{\delta_2^2+\alpha_2^2}$, is the phase-sensitive parametric pump driving term. Here $\psi = -\arctan(\delta_2/\alpha_2)$ is the phase offset between the cw pump field $B_{in}$ and the signal field $A$. To maximize the parametric pump driving term, one must choose the wave-vector mismatch parameter that equals this phase offset, namely $\xi = \psi$.

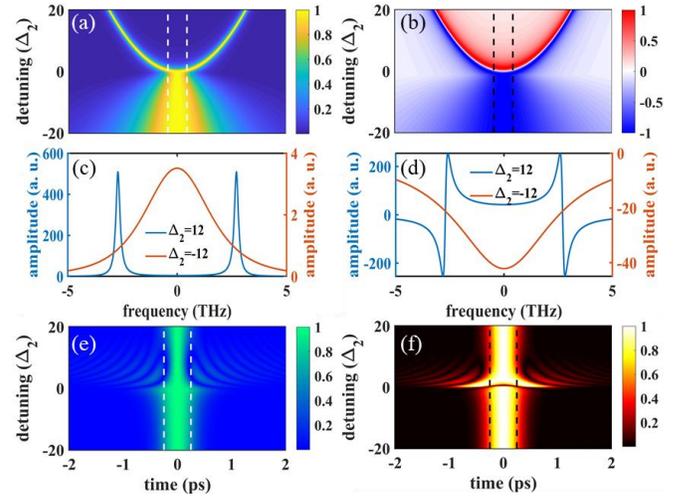

Fig. 2. Effect of normalized pump-resonance phase detuning $\Delta_2$ on the (a) dispersive effective TPA $X(\Omega)$ and (b) dispersive effective Kerr effect $Y(\Omega)$. The dashed (white and black) lines in (a) and (b) show the bandwidth of the test 500-fs Gaussian pulse. Line profiles of $X(\Omega)$ and $Y(\Omega)$ at $\Delta_2 = \pm 12$ are plotted in (c) and (d), respectively. Effect of $\Delta_2$ on the convolution of test Gaussian pulse and inverse Fourier transformation of the $X(\Omega)$ and $Y(\Omega)$ are shown in (e) and (f), respectively. The dashed (white and black) lines in (e) and (f) show the duration of the test 500-fs Gaussian pulse. $D_2 = 4.15 \times 10^4$ fs².

To shed light on the frequency-dependent nonlinear response function, we divide $J(\Omega) = X(\Omega) - iY(\Omega)$ into the real and imaginary parts to examine them independently:

$$X(\Omega) = \frac{1}{\alpha_2} \frac{1}{1+\left(\Delta_2 - D_1\Omega - D_2\Omega^2\right)^2}, \quad (7)$$

$$Y(\Omega) = \frac{1}{\alpha_2} \frac{\Delta_2 - D_1\Omega - D_2\Omega^2}{1+\left(\Delta_2 - D_1\Omega - D_2\Omega^2\right)^2}. \quad (8)$$

Here $\Delta_2 = \delta_2/\alpha_2$ is the normalized pump-resonance phase detuning, $D_1 = \Delta k^{'} L/\alpha_2$ is the normalized temporal walk-off and $D_2 = k_2^{"}L/(2\alpha_2)$ is the normalized pump group delay dispersion (GDD). $X(\Omega)$ and $Y(\Omega)$ resemble the dispersive TPA and the dispersive Kerr effect respectively. Importantly, both $X(\Omega)$ and $Y(\Omega)$ are set solely by the design of the pump field parameters. We first consider the case of zero GVM and then analyze the temporal walk-off as a perturbation to the quadratic soliton.

Figure 2 plots the dispersive effective third-order nonlinearity as a function of normalized pump-resonance phase detuning $\Delta_2$. Similarly,

Fig. S1 (see Supplement 1, section 1) plots the dispersive effective third-order nonlinearity as a function of normalized pump GDD $D_2$. Two distinct regimes can be evidently identified and divided into the upper branch where $\Delta_2 D_2 > 0$ and the lower branch where $\Delta_2 D_2 < 0$ (Figs. 2(a) and 2(b)). The upper branch is characterized by the two resonance-enhanced effective TPA peaks and associated nonlinear phase anomalies symmetrically located at $\Omega = \pm\sqrt{\Delta_2/D_2}$. To elucidate their effect on a pulse, a Gaussian signal field with a transform limited pulse duration (full width at half maximum, FWHM) $\Delta T$ of 500 fs is introduced to convolve with the inverse Fourier transformation of the nonlinear response function (Figs. 2(e) and 2(f)). Long oscillatory tails resulting from the narrowband perturbation are excited by the main pulse, similar to the effect of mode-crossing induced perturbation to the dissipative Kerr soliton and Kerr frequency comb generation in cw-pumped cubic nonlinear cavities [27-30]. In general, the effect leads to disruptive spectral modulation and single pulse destabilization due to the long-range interaction nature of such long oscillatory tails. A straightforward strategy to avoid the detrimental effect is to keep the two resonant peaks far separated by more than the pulse bandwidth, namely $2\sqrt{\Delta_2/D_2} \gg \pi/\Delta T$, through the choice of small pump GDD and large pump-resonance phase detuning. In the lower branch, no such narrowband resonance phenomenon is observed and the frequency response shows a smooth inverse bandpass filtering behavior with the bandwidth approaching $2\sqrt{(\sqrt{2}-1)|\Delta_2/D_2|}$ when $|\Delta_2|$ is large. The smooth profiles of $X(\Omega)$ and $Y(\Omega)$ leads to a much cleaner pulse shape in the time domain and the inverse bandpass filtering can partially compensate for the regular spectral roll-off resulting in some bandwidth enhancement.

From Eq. (7) and Eq. (8), $Y(\Omega=0)/X(\Omega=0)$ is proportional to the normalized pump-resonance detuning $\Delta_2$ and thus the effective Kerr effect $Y(\Omega)$ will dominate over the effective TPA effect $X(\Omega)$ at large normalized pump-resonance detuning when $|\Delta_2| \gg 1$. Eq. (5) can then be further simplified to the form of parametrically driven NLSE [23, 24]:

$$t_R \frac{\partial A}{\partial t} = \left(-\alpha_1 - i\delta_1 - i\frac{k_1^{"}L}{2}\frac{\partial^2}{\partial \tau^2}\right)A + i\gamma_{eff}L|A|^2 A + i\rho A^*, \quad (9)$$

where $\rho \approx \kappa L \mathrm{sinc}(\pi/2)\sqrt{\theta_2}B_{in}/\sqrt{\delta_2^2 + \alpha_2^2}$ is the parametric drive coefficient and $\gamma_{eff} \approx \kappa^2 L \mathrm{sinc}^2(\pi/2)/\delta_2$ is the effective Kerr nonlinear coefficient. Importantly and uniquely, the effective Kerr nonlinearity depends on not only the second-order nonlinearity coupling coefficient $\kappa$ but also the medium length $L$ and the pump-resonance detuning $\delta_2$. Of note, sign of the effective Kerr nonlinearity is solely determined by the choice of pump-resonance detuning such that bright quadratic solitons can exist in both normal and anomalous dispersion regimes as long as $\delta_2 k_1^{"} < 0$. In addition, as shown in Fig. 2, dispersion characteristics of the effective Kerr nonlinearity that perturb the quadratic soliton are divided into the upper and lower branches, depending on the choice of the pump GVD and pump-resonance detuning. To sum up, upper branch quadratic solitons occur in the parameter space where $\delta_2 k_1^{"} < 0$ and $k_1^{"}k_2^{"} < 0$ while lower branch quadratic solitons exist in the parameter space where $\delta_2 k_1^{"} < 0$ and $k_1^{"}k_2^{"} > 0$ (Table 1).

By solving Eqs. (1)-(4) with the standard split-step Fourier method, pulse shapes and optical spectra representative of the upper branch (first row in Table 1) and the lower branch (last row in Table 1) quadratic solitons are shown in Fig. 3. In both branches, terahertz comb bandwidth and femtosecond pulse duration are attainable. Furthermore, there is no cw spectral peak and temporal background on the DOPO signal output. As described in the following, there are a few apparent differences between the upper and lower branch quadratic solitons resulting from the distinct dispersion characteristics of their effective TPA and Kerr nonlinearity (Fig. 2). First, the lower branch quadratic soliton requires a lower pump power due to the overall lower effective TPA loss. Second, narrowband resonance-enhanced perturbation in the upper branch manifests itself into the local modulation of both the pump and the signal spectra as expected. However, quite surprisingly, the signal pulse shape remains clean and only the pump acquires the long oscillatory tails in the DOPO process. The pump oscillatory tail renders itself into the signal temporal phase oscillation without causing much distortion to the signal pulse shape. In the lower branch, the quadratic soliton exhibits not only clean pulse shape without any oscillatory tail but also smooth comb spectrum without any local modulation. Furthermore, the 10-dB bandwidth is >15 % larger as expected from the inverse bandpass filtering effect in the lower branch. Finally, while the lower branch quadratic soliton remains close to its transform limit, the upper branch quadratic soliton exhibits an appreciable linear frequency chirp within its FWHM pulse duration due to the signal temporal phase oscillation.

**Table 1. Existence of quadratic soliton with respect to the signs of the pump-resonance detuning, signal GVD and pump GVD.**

| quadratic soliton type | $\delta_2$ | $k_1^{"}$ | $k_2^{"}$ |
|---|---|---|---|
| upper branch | + | − | + |
| upper branch | − | + | − |
| lower branch | + | − | − |
| lower branch | − | + | + |

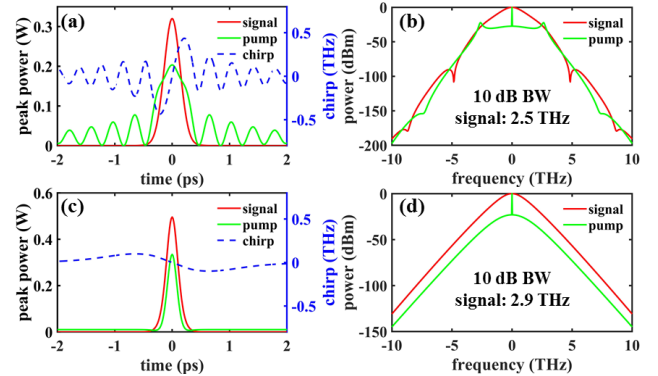

Fig. 3. Pulse shape and optical spectrum of the upper branch quadratic soliton (a)(b) and the lower branch quadratic soliton (c)(d) in the GV matched DOPO. $\alpha_1 = \alpha_2 = \theta_1 = \theta_2 = \pi/1600$, $L = 1$ mm, $A_{eff} = 28$ μm², $\kappa = 54.2$ W$^{-1/2}$m$^{-1}$, $D_2 = 4.15\times10^4$ fs². (a)(b) $|B_{in}|^2 = 5$ mW, $\Delta_2 = 2\Delta_1 = 12$, $k_1^{"} = -325$ fs²/mm. The 10-dB comb bandwidth is 2.5 THz and the FWHM pulse duration is 341 fs, positively chirped from its transform limit of 282 fs. (c)(d) $|B_{in}|^2 = 3$ mW, $\Delta_2 = 2\Delta_1 = -12$, $k_1^{"} = 325$ fs²/mm. The 10-dB comb bandwidth is 2.9 THz and the FWHM pulse duration is 252 fs. The negative chirp has negligible effect on the pulse duration.

The parameters used for the upper branch quadratic soliton (Figs. 3(a) and 3(b)) can be readily achieved in a monolithic PPLN waveguide microresonator with a 28-μm² mode area, a 1-mm cavity length, a 1262-nm cw pump wavelength, and a 2524-nm signal center wavelength. Due to the large effective Kerr nonlinear coefficient of 79 W$^{-1}$m$^{-1}$, high-quality quadratic soliton can be obtained with a cw pump power as low as 5 mW. It is more challenging to fulfill the existence condition of lower branch quadratic soliton, $\delta_2 k_1^{"} < 0$ and $k_1^{"}k_2^{"} > 0$, in conventional bulk materials and waveguide designs. Multiple zero-dispersion points between the pump and signal wavelengths are required to meet the existence condition and a strategy to achieve it is by employing a recently studied sandwich waveguide structure [31]. Besides the PPLN, other commonly used MIR nonlinear materials such as $CdSiP_2$, $ZnGeP_2$, OP-GaP, and OP-GaAs are all potential platforms for the quadratic soliton mode-locked DOPO, extending the signal wavelengths to cover most of the MIR spectral region (3-10 μm, see Supplement 1, section 2). In addition, the cavity can be pumped at the FF wavelength and configured for the doubly resonant SHG cavities as discussed in the Supplement 1, section 3. Here the governing single mean-field equation has the same form with the Lugiato-Lefever equation [22], and thus quadratic soliton also exists with the characteristics resembling dissipative Kerr soliton and Kerr frequency comb generation in cw-pumped cubic nonlinear cavities.

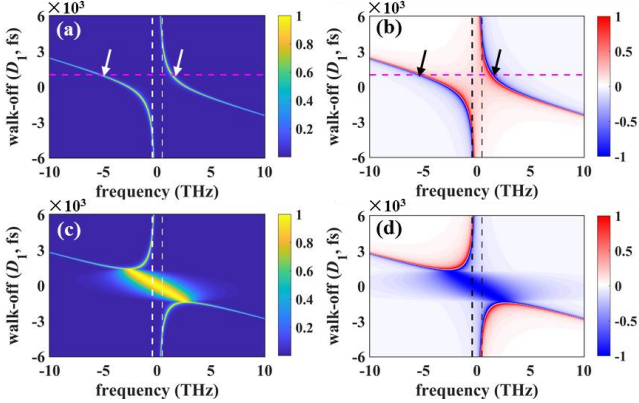

Fig. 4. Effect of normalized temporal walk-off $D_1$ on the dispersive effective TPA $X(\Omega)$ and dispersive effective Kerr effect $Y(\Omega)$ in the upper branch where $\Delta_2 = 12$ (a)(b) and lower branch where $\Delta_2 = -12$ (c)(d). The dashed (white and black) lines show the bandwidth of the test 500-fs Gaussian pulse. The arrows in (a) and (b) indicate the spectral locations of the resonant TPA peaks and the associated nonlinear phase anomalies. $D_2 = 4.15 \times 10^4$ $fs^2$.

When the normalized temporal walk-off $D_1$ is considered, the frequency-dependent nonlinear response function $J(\Omega)$ is evidently perturbed and asymmetry occurs in both $X(\Omega)$ and $Y(\Omega)$ as shown in Fig. 4. In the upper branch where $\Delta_2 D_2 > 0$, real roots of $\Delta_2 - D_1\Omega - D_2\Omega^2 = 0$ can be found and thus two resonance-enhanced effective TPA peaks and associated nonlinear phase anomalies always exist (Figs. 4(a) and 4(b)). As the temporal walk-off $D_1$ increases, the resonance closer to the center frequency ($\Omega = 0$) asymptotically approaches $\Omega = \Delta_2/D_1$ while the other resonance continues to linearly move away from the center frequency. In the lower branch where $\Delta_2 D_2 < 0$, the behavior of the frequency-dependent nonlinear response function is divided into two distinct regimes (Figs. 4(c) and 4(d)). When the temporal walk-off is small such that $|D_1| < \sqrt{-4\Delta_2 D_2}$, real root of $\Delta_2 - D_1\Omega - D_2\Omega^2 = 0$ does not exist and thus no narrowband resonance phenomena is present. In this regime, the smooth profiles of $X(\Omega)$ and $Y(\Omega)$ guarantee the GVM has minimal perturbative effect to the quadratic soliton. On the other hand, resonance-enhanced effective TPA peaks and associated nonlinear phase anomalies reappear as the temporal walk-off increases above $\sqrt{-4\Delta_2 D_2}$. Similarly, the resonance closer to the center frequency asymptotically approaches $\Omega = \Delta_2/D_1$.

A straightforward strategy to avoid the detrimental narrowband perturbation is to keep the closer resonance well away from the center frequency by more than the pulse bandwidth, namely $2|\Delta_2/D_1| \gg \pi/\Delta T$, through the choice of small normalized temporal walk-off and large normalized pump-resonance phase detuning. While $\Delta_2$ is a readily accessible parameter in the experiment, increasing it to compensate for the GVM perturbation is not a suggested way because a higher $\Delta_2$ not only reduces the cavity enhancement but also lowers the coefficients of the parametric drive and the effective Kerr nonlinearity (see Eq. (9)), resulting in a significantly elevated pump power requirement. In contrast, lowering the normalized temporal walk-off by increasing the pump cavity loss is a better strategy to compensate for the GVM perturbation as shown in Fig. 5.

Figures 5(a) and 5(b) plot the pulse shape and optical spectrum of an upper branch quadratic soliton under a GVM perturbation of $\Delta k' = 2$ fs/mm and $D_1 = 10^3$ fs. Asymmetry is evidently observed in both the pulse shape and the optical spectrum. The two spectral peaks of the pump are located precisely at the points indicated by the arrows in Figs. 4(a) and 4(b). The temporal walk-off of the pump selectively enhances the oscillatory tails on either side of the pump pulse, depending on the sign of $D_1$, but the signal pulse shape remains clean and minimally perturbed. When the GVM $\Delta k'$ further increases to 10 fs/mm, the oscillation becomes so severe that even the signal breaks up into multiple pulses and the quadratic soliton mode-locking ceases. As discussed in the previous paragraph, a strategy to reestablish the quadratic soliton mode-locking is increasing the total pump linear cavity loss $\alpha_2$. As shown in Figs. 5(c) and 5(d), stable quadratic soliton can persist under an even larger GVM perturbation of $\Delta k' = 20$ fs/mm as long as $\alpha_2$ is linearly increased such that the normalized temporal walk-off $D_1$ remains unchanged. Compared with the results shown in Fig. 5(a) and 5(b), the quadratic soliton shown in Fig. 5(c) and 5(d) drifts at a higher speed due to a larger GVM but the pump spectral peaks are at the identical locations due to the same normalized temporal walk-off. The greater tolerance over the GVM perturbation comes at the price of higher pump power requirement linearly scaled with $\alpha_2$.

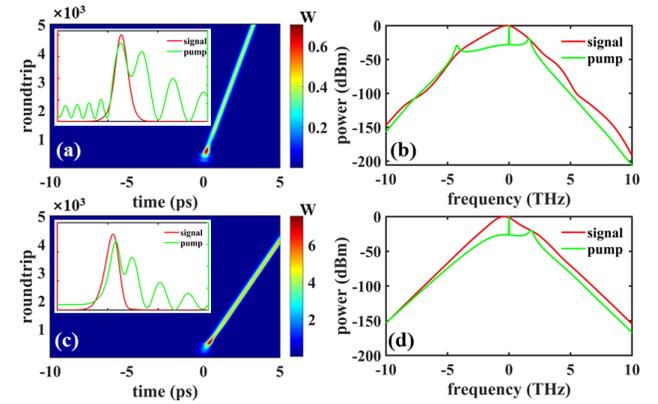

Fig. 5. Effect of normalized temporal walk-off $D_1$ on (a)(c) the pulse shapes and (b)(d) the optical spectra of the upper branch quadratic solitons. (a)(b) Parameters are the same as those used in Fig. 3(a), except for the pump field $|B_{in}|^2 = 7$ mW and the normalized temporal walk-off $D_1 = 10^3$ fs. (c)(d) $\alpha_1 = \theta_1 = \pi/1600$, $\alpha_2 = \theta_2 = \pi/160$, $\Delta_2 = 2\Delta_1 = 12$, $|B_{in}|^2 = 70$ mW, $D_1 = 10^3$ fs, and $D_2 = 4.15 \times 10^3$ $fs^2$.

## 3. CONCLUSION

In conclusion, we study the previously unexplored parameter space and unveil the existence condition of high-quality passively mode-locked DOPO. We identify the similarities between the governing equation of the DOPO dynamics and the parametrically driven NLSE. Such an analogy facilitates our investigation into the nature of the quadratic soliton in cw-pumped quadratic nonlinear cavities. Sign of the effective Kerr nonlinearity is solely determined by the choice of pump-resonance detuning such that bright quadratic solitons can exist in both normal and anomalous dispersion regimes as long as the signal GDD has an opposite sign. The dominant perturbation to the quadratic soliton results from the dispersion of the effective third-order nonlinearity; its characteristics can be divided into two distinct branches depending on the sign of the multiplication of the pump GDD and the signal GDD. In the absence of temporal walk-off, such intrinsic perturbation to the quadratic soliton can be minimized through the choice of small pump

GDD and large pump-resonance phase detuning. When the temporal walk-off is present, the dispersion of the effective third-order nonlinearity becomes highly asymmetric and the recommended strategy to alleviate the additional GVM perturbation is to increase the total pump linear cavity loss at the cost of higher pump power requirement. Two examples, one with a PPLN waveguide microresonator and one with an OP-GaAs free-space cavity, are given to demonstrate the feasibility of generating terahertz comb bandwidth and femtosecond pulse duration. The quadratic soliton mode-locking principle can be further extended to other material platforms, making it a competitive ultrashort pulse and broadband comb source architecture at the MIR spectral region.

**Funding.** This work has been supported by the Office of Naval Research (ONR) under award number N00014-19-1-2251.

**Disclosures.** The authors declare no conflicts of interest.

**Acknowledgment**. We thank Prof. Changdong Chen, Dr. Kunpeng Jia, Dr. Bowen Li and Dr. Xiaohan Wang for fruitful discussion.

See Supplement 1 for supporting content.